\documentclass{aa}

\usepackage{graphicx}
\usepackage{epstopdf}
\usepackage{txfonts}

\usepackage{xcolor}                       
\usepackage[normalem]{ulem}     

\bibpunct{(}{)}{;}{a}{}{,} 

\begin{document}

\title{Oscillation of solar radio emission at coronal acoustic cut-off frequency}

\author{
                Pylaev, O. S.\inst{1,2}
        \and
                Zaqarashvili, T. V. \inst{1,3,5}
        \and
                Brazhenko, A.I.\inst{2}
        \and
                Melnik, V. N.\inst{4}
        \and
                Hanslmeier, A.\inst{3}
                \and
                Panchenko, M.\inst{1}
}

\institute{
                Space Research Institute, Austrian Academy of Sciences, Schmiedlstrasse 6, 8042 Graz, Austria
        \and
                Poltava Gravimetrical Observatory at Institute of Geophysics, National Academy of Sciences of Ukraine, Myasoedova str., 27/29, 36014, Poltava, Ukraine
        \and
                IGAM, Institute of Physics, University of Graz, Universit\"atsplatz 5, 8010 Graz, Austria
                \email{teimuraz.zaqarashvili@uni-graz.at}
        \and
                Institute of Radio Astronomy, National Academy of Sciences of Ukraine, Chervonopraporna st., 4, 61002, Kharkiv, Ukraine
        \and
                Abastumani Astrophysical Observatory at Ilia State University, Kakutsa Cholokashvili Ave 3/5, 0162 Tbilisi, Georgia
}

\date{Received <date> /
                Accepted <date>
}

\abstract{
Recent SECCHI COR2 observations on board STEREO-A spacecraft have detected density structures at a distance of 2.5--15~$R_0$ propagating with periodicity of about 90~minutes. The observations show that the density structures  probably formed in the lower corona.
We used the large Ukrainian radio telescope URAN-2 to observe type IV radio bursts in the frequency range of 8--32~MHz during the time interval of 08:15--11:00~UT on August 1, 2011. Radio emission in this frequency range  originated at the distance of 1.5--2.5 $R_0$ according to the Baumbach-Allen density model of the solar corona.
Morlet wavelet analysis showed the periodicity of $\sim$80~min in radio emission intensity at all frequencies, which demonstrates that there are quasi-periodic variations of coronal density at all heights.
The observed periodicity corresponds to the acoustic cut-off frequency of stratified corona at a temperature of 1~MK.
We suggest that continuous perturbations of the coronal base in the form of jets/explosive events generate acoustic pulses, which propagate upwards and leave the wake behind oscillating at the coronal cut-off frequency. This wake may transform into recurrent shocks due to the density decrease with height, which leads to the observed periodicity in the radio emission. The recurrent shocks may trigger quasi-periodic magnetic reconnection in helmet streamers, where the opposite field lines  merge and consequently may generate periodic density structures observed in the solar wind.  
}


\maketitle

\titlerunning{<titlerunning>}
\authorrunning{<authorrunning>}

\section{Introduction}
        \label{s_Introduction}

Oscillations and waves are ubiquitous along the whole solar atmosphere. The photosphere is dominated by 5 min oscillations, which are global standing acoustic modes \citep{Gizon2005}. The dominant oscillation period in the solar chromosphere is 3 min oscillations, which correspond to the chromospheric acoustic cut-off period \citep{Fleck1991,Kuridze2009,Sych2012}. Waves and oscillations have been continuously observed in different magnetic structures: in sunspots \citep{khomenko2015}, in spicules \citep{Zaqarashvili2009}, and in coronal loops \citep{nakariakov2005}. Waves are important for various reasons. They may lead to turbulence and heating in the ambient plasma, and can also  be used to estimate plasma parameters in different layers and magnetic structures.

In situ measurements of solar wind density near the Earth show periodic fluctuations on timescales of several to tens of minutes \citep{Viall2009}. The formation mechanism of these fluctuations is not yet clear. The authors have suggested that the observed fluctuations are not excited in situ in the interplanetary medium, but are transported from the solar corona by the solar wind. \citet{Viall2010} observed similar periodic density structures as streamer blobs using white-light images of the solar wind taken with the Heliospheric Imager \citep{Eyles2009} at the Sun Earth Connection Coronal and Heliospheric Investigation (SECCHI) \citep{Howard2008} on board the Solar Terrestrial Relations Observatory Ahead (STEREO-A) spacecraft \citep{Kaiser2008}. They detected  multiple periodicity, the shortest being about 100~min. The authors concluded that the density structures are formed below 15~$R_0$, where $R_0$ is the solar radius, and then propagate outwards. In order to study the physical processes involved in the formation of the periodic density structures observed in the solar wind, \citet{Viall2015} have performed a similar analysis of the images taken at the distance 2.5--15~$R_0$ with SECCHI COR2. They found that the density structures occur as blobs with a periodicity of about 90~minutes near coronal streamers. The authors did not find any evidence of the formation of the fluctuations at these heights and thus  concluded that the formation could happen in the lower corona. Several models involving magnetic reconnection in formation of the periodic density structures have been suggested \citep{Antiochos2011,Linker2011,Titov2011,Lynch2014}, but  none could  fully explain the observed periodicity. {\citet{Allred2015} were able through numerical simulations to generate quasi-periodic helmet stream disconnections with the period of 2 hours; however, the location of the first reconnection event was estimated to be near 6~$R_0$, which is far away from the regions where density structures start to be observed (2.5~$R_0$).

\begin{figure*}
\centering
\includegraphics[width=17cm]{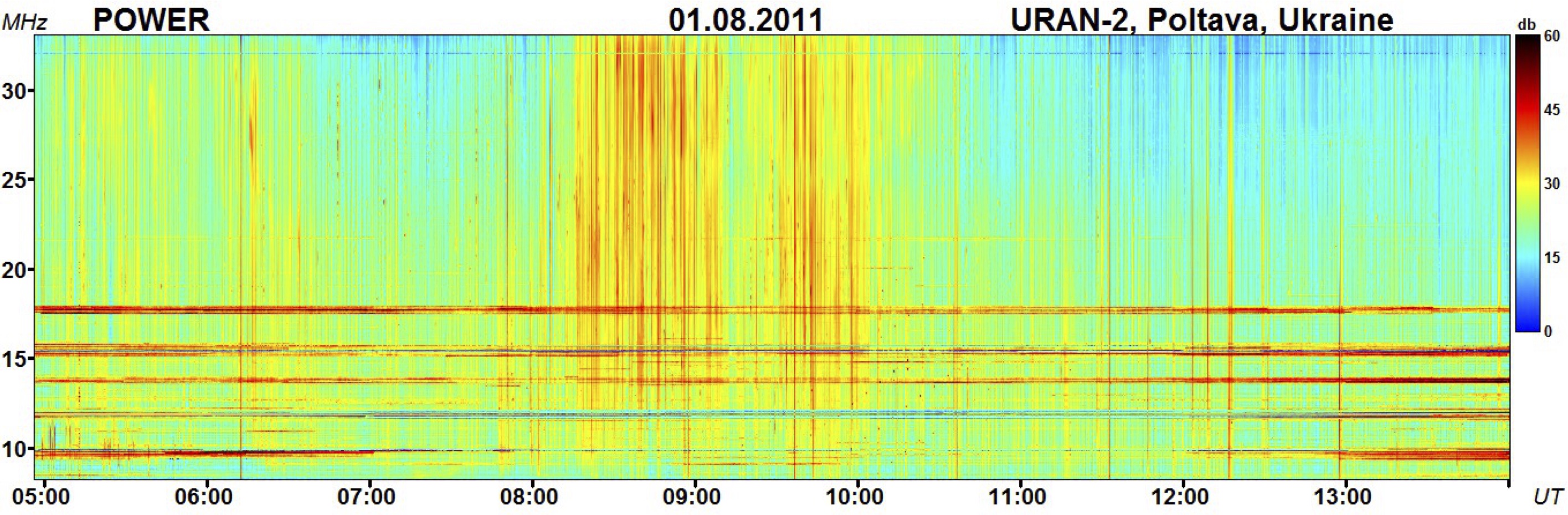}
\caption{Dynamic spectrum of solar radio emission observed by radio telescope URAN-2 on August 1, 2011. }
\label{fig_dynamic_spectrum}
\end{figure*}

We  analysed radio emission at the frequencies of 8--32~MHz, which correspond to a distance of about 1.5--2.5~$R_0$, obtained with radio telescope URAN--2 on~August~1, 2011. The radio observation allowed us to search for the periodicity source at heights lower than the COR2 FOV. We analysed the intensity of radio emission during type~IV~bursts and searched for the periodicity using the wavelet tool. We show that a periodicity of about 80~minutes is seen at the observed distances, which indicates that the source of the oscillation is located in the lower corona. We suggest that the periodic density structures observed by \citet{Viall2015} can be explained by oscillations at coronal acoustic cut-off frequency.

\section{Observations}
        \label{s_Observations}

The radio telescope URAN-2 is located near Poltava (Ukraine) and it operates in the frequency range of 8--32~MHz. The effective area of the telescope is about 28000~m$^2$. The angular beam size at 25~MHz is about 7$^0$--3.5$^0$. The sensitivity in  single operation mode is about 500~Jy. Elements of the antenna array in a form of cross dipole enable polarization measurements of the radio emission \citep{Megn2003, Brazhenko2005}. The digital spectrograph DSPz \citep{Ryabov2010}, which is used at the radio telescope, can register radio emission with time and frequency resolution of about 10~ms and 4~kHz, respectively.

During 04:56--13:59~UT on August 1, 2011, URAN-2 registered solar radio emission in frequency range of 8--32~MHz with maximum frequency resolution. A time resolution of about 100~ms was used. A type IV burst was detected in the frequency range of 10--32~MHz. It started at $\sim$~08:15~UT and ended at $\sim$~11:00~UT. The burst was probably connected either with a C1.4-type flare which occurred during 08:10-08:17 UT in the active region NOAA 11263 or with a C4-type flare which occurred during 07:10-07:44 UT in the active region NOAA 11261. The polarization of the type IV burst was about 40--50~\%. This event had a marked fine structure, which has been discussed  by \citet{Antonov2014}.  A type III storm was registered before this type IV burst, which was probably connected with the C4 flare. Moreover, after the type IV burst another type III storm was seen. Figure~\ref{fig_dynamic_spectrum} shows the dynamic spectrum of solar radio emission observed by radio telescope URAN-2 on August 1, 2011.

Decameter type III and IV bursts are usually linked with solar flares and explained by a plasma emission mechanism: fast energetic electrons accelerated during a flare generate Langmuir waves in coronal loops, which then lead to the emission of electromagnetic waves. The frequency of radio emission equals the local plasma frequency or double the local frequency; therefore, it corresponds to the value of the local density. Type III bursts are short radio bursts with very strong frequency drift rates, which indicates  their connection with fast electron beams propagating along open magnetic field lines. On the other hand, type IV radio bursts are probably connected with trapped electrons in closed magnetic field lines of post-flare arcades. Type IV bursts are usually accompanied by coronal mass ejections (CMEs), but sometimes there are no visible CMEs during type IV bursts. The type IV burst discussed in this paper was not accompanied by visible CME.}

\section{Data analysis}
        \label{s_Observations}

\begin{figure*}
\centering
\includegraphics[width=17cm]{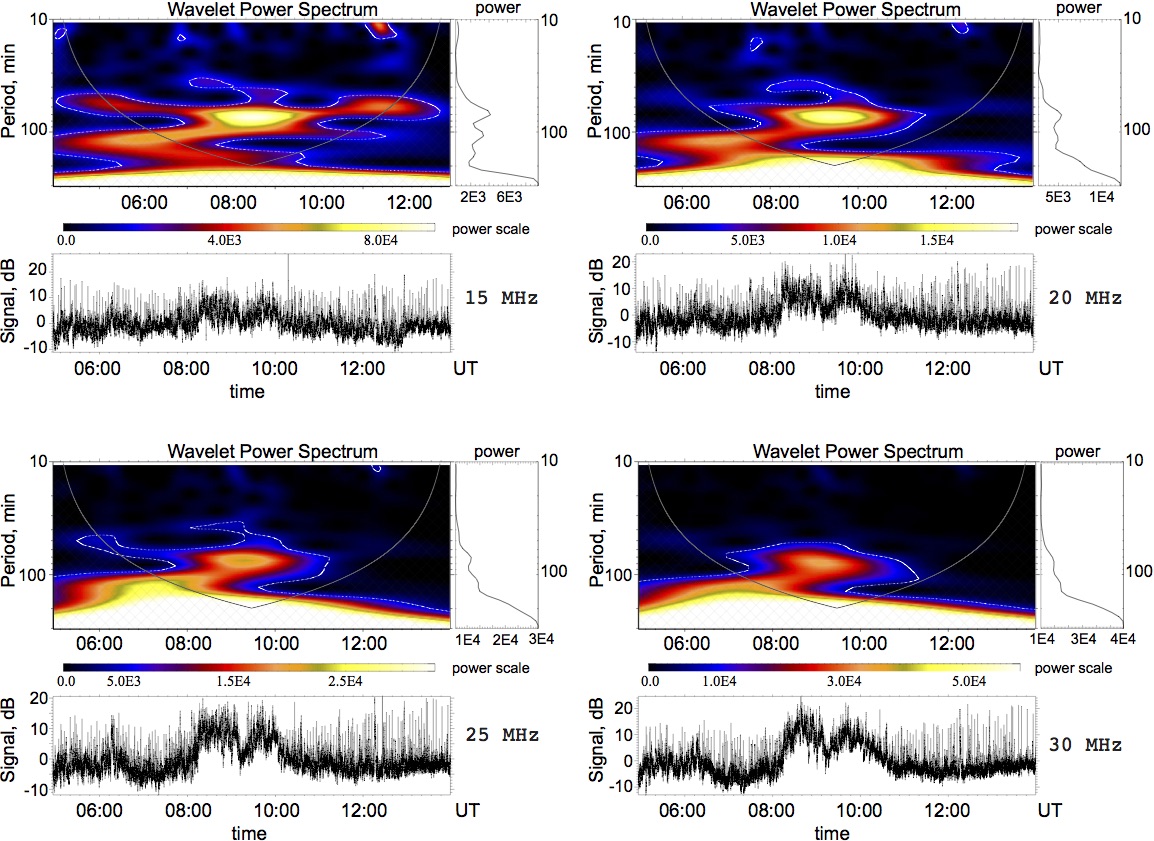}
\caption{Wavelet power spectrum of radio emission intensity at 15~MHz (upper left panel), 20~MHz (upper right panel), 25~MHz (lower left panel), and 30~MHz (lower right panel) during 5:00--14:00~UT.}
\label{fig_wavelet}
\end{figure*}

We use a wavelet analysis to find oscillations of the radio emission intensity and to define their temporal behaviour  at four different frequencies: 15~MHz, 20~MHz, 25~MHz, and 30~MHz (see Figure~\ref{fig_wavelet}). The most powerful period on the wavelet power spectrum at all frequencies is located at about 77$\pm$7~min. It starts with type IV bursts and continues for about 2--2.5~periods. The shorter period oscillations are also seen on the power spectrum, but with significantly smaller amplitude. The oscillations are probably connected to the kink oscillations of coronal loops  \citep{Zaqarashvili2013} and so here we  concentrate only on the long-period oscillations. \citet{Ireland2015} have shown that the power spectrum of coronal emission intensity is a power law even in the low corona; therefore, the appropriate way to identify discrete signals is to assume a red noise background spectrum. In order to check the reality of oscillations in our data, we assumed a red noise background spectrum in the wavelet analysis following  \citet{Torrence1998}, which still shows strong power at 77~min oscillations. Therefore, the oscillations are due to real discrete acoustic waves and do not belong to turbulent random processes.
 
The oscillation period is almost the same at all frequencies of type IV radio bursts, hence it does not change significantly with height. Type IV radio bursts are generally produced by an emission mechanism near the plasma frequency, which depends on the electron density $n_e$ as
\begin{equation}\label{plasma-frequency}
\nu_p= 8980 \sqrt{n_e},
\end{equation}
where $\nu_p$ is in Hz and $n_e$ is in cm$^{-3}$.  Therefore, the radio emission at a particular frequency is excited at a particular local density of emitted plasma. The density of the solar corona generally decreases with distance, hence the radio emission at different frequencies corresponds to different heights above the solar surface.
The density structure of the quiet Sun corona can be approximated by several different models \citep{Allen1947,Newkirk1961,Mann1999} which give different excitation heights of radio frequencies. We use the Baumbach-Allen model, which implies that the radio emission in the frequency range of 10--32~MHz  originated somewhere between the distances 1.5 and 2.5~$R_0$. 

The periodicity of $\sim$80 min found in radio emission is near the period reported by \citet{Viall2015} indicating that the  phenomena may both have the same source, which is located at least below 2.5~$R_0$.
The periodicity is in the range of the acoustic cut-off period of gravitationally stratified corona \citep{Roberts2004,Afanasyev2015}. All of this  suggests that the periodicity is caused by the oscillation of the coronal plasma at an acoustic cut-off frequency similar to chromospheric 3 min oscillations. In the next section, we study this problem using hydrodynamic equations for stratified medium.

\section{Oscillation of coronal plasma at acoustic cut-off frequency behind a propagating pulse}

The equilibrium density in the stratified corona with homogeneous temperature $T=const$ is expressed as
\begin{equation}
\rho(z)=\rho_0e^{-{R_0\over H_n}{z\over {R_0+z}}},
\label{eq:S1}
\end{equation}
where $H_n=2kTR^2_0/G M m$ is the pressure scale height near the solar surface, $z$ is the vertical coordinate, $R_0$ is the solar radius, $G=6.674 \cdot 10^{-8}$ cm$^3$ g$^{-1}$ s$^{-2}$ is the gravitational constant, $M=1.989\cdot 10^{33}$ g is the solar mass, $k=1.38\cdot 10^{-16}$ erg K$^{-1}$ is the Boltzmann constant, and $m=1.67 \cdot 10^{-24}$ g is the proton mass. It differs from the above-mentioned density models, but due to the simplicity of analytical presentation    this expression can be used.

\begin{figure}%
\includegraphics[width=19cm]{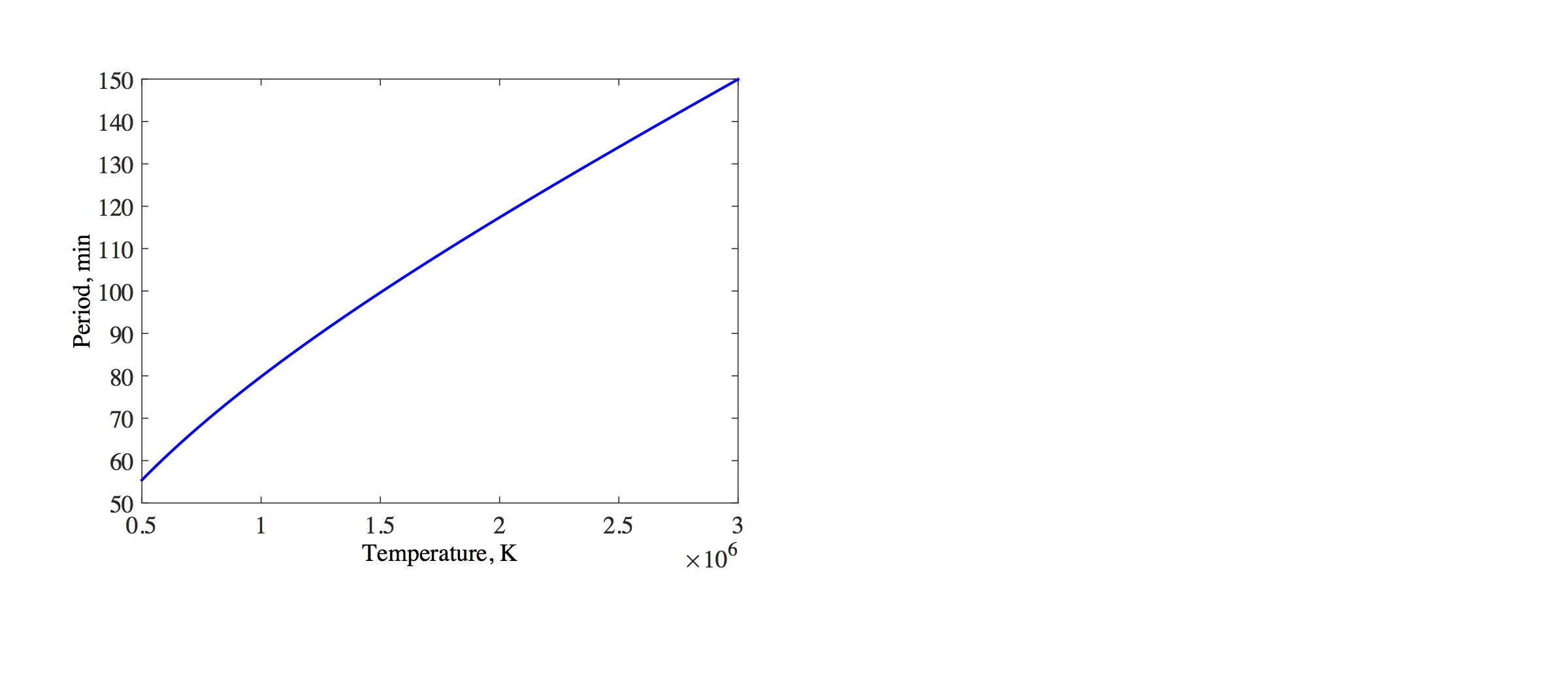}
\caption{Acoustic cut-off period vs plasma electron temperature in the solar corona. }
\label{cut-off}
\end{figure}%

The ideal linear hydrodynamic equations lead to the  second-order differential equation
\begin{equation}
\label{4}
{{\partial^2 u_z}\over {\partial z^2}} - {{\gamma g}\over {c^2_s}}{{\partial u_z}\over {\partial z}} - {{1}\over {c^2_s}}{{\partial g}\over {\partial z}}u_z-{{1}\over {c^2_s}}{{\partial^2 u_z}\over {\partial t^2}}=0,
\end{equation}
where $u_z$ is the perturbation of vertical velocity, $g={{GM}/{(R_0+z)^2}}$ is the gravitational acceleration, and $c_s=\sqrt{2 \gamma k T/m}$ is the adiabatic sound speed.

Using the substitution of $u_z=Q(t,z)\exp{\left (\int{(\gamma g/2c^2_s)dz}\right )}$ we obtain the Klein-Gordon equation \citep{Rae1982, Spruit1983, Roberts2004, Zaqarashvili2008, Kuridze2009,Afanasyev2015}
\begin{equation}
\label{5}
{{\partial^2 Q}\over {\partial t^2}}-c^2_s{{\partial^2 Q}\over {\partial z^2}} + \Omega^2_sQ=0,
\end{equation}
where
\begin{equation}
\label{6}
\Omega^2_s(z)={{\gamma^2 g^2(z)}\over {4c^2_s}} + {{2-\gamma}\over {2}}{{\partial g}\over {\partial z}}.
\end{equation}
This equation describes the vertically propagating acoustic waves in the stratified atmosphere. We note that the last term in Eq.~(\ref{6}) corresponds to the variation of gravitational acceleration with height, and that it is absent in previous studies. The waves with the lower frequency than the cut-off frequency are evanescent. The cut-off frequency for the lower corona can be estimated from Eq.~(\ref{6}) assuming $z \ll R_0$:
\begin{equation}
\label{7}
\Omega_{c}=\sqrt{{{\gamma^2 G^2 M^2}\over {4c^2_sR^4_0}} -{{(2-\gamma)GM}\over {R^3_0}}}.
\end{equation}

Figure~\ref{cut-off} shows the dependence of acoustic cut-off period, $T_c=2\pi/\Omega_{c}$, on plasma temperature. The temperature of $10^6$~K yields a period of 80~min, which agrees  very well with the periodicity found in our data. This means that acoustic waves with periods of $> 80$~min are evanescent in the lower corona. On the other hand, the lower corona is very dynamic with many observed jets, local transients, etc., originating near the chromosphere/transition region. Micro and nanoflares also lead to impulsive energy release in the lower corona. These impulsive events may generate acoustic pulses, which propagate from the lower corona upwards. In order to study the dynamics of the corona after the propagation of pulses, we use simple impulsive force in time and in coordinate, then  Eq.~(\ref{5}) can be rewritten as
\begin{equation}
\label{8}
{{\partial^2 Q}\over {\partial z^2}}-{{\Omega^2_c}\over {c^2_s}}Q-{{1}\over {c^2_s}}{{\partial^2 Q}\over {\partial t^2}}=-4\pi\delta(t)\delta(z),
\end{equation}
where $z>-\infty$, $t>0$ and the pulse is set at $t=0$ and $z=0$, which coincides to the lower boundary of the corona.

\begin{figure}%
\centering
\includegraphics[width=10cm]{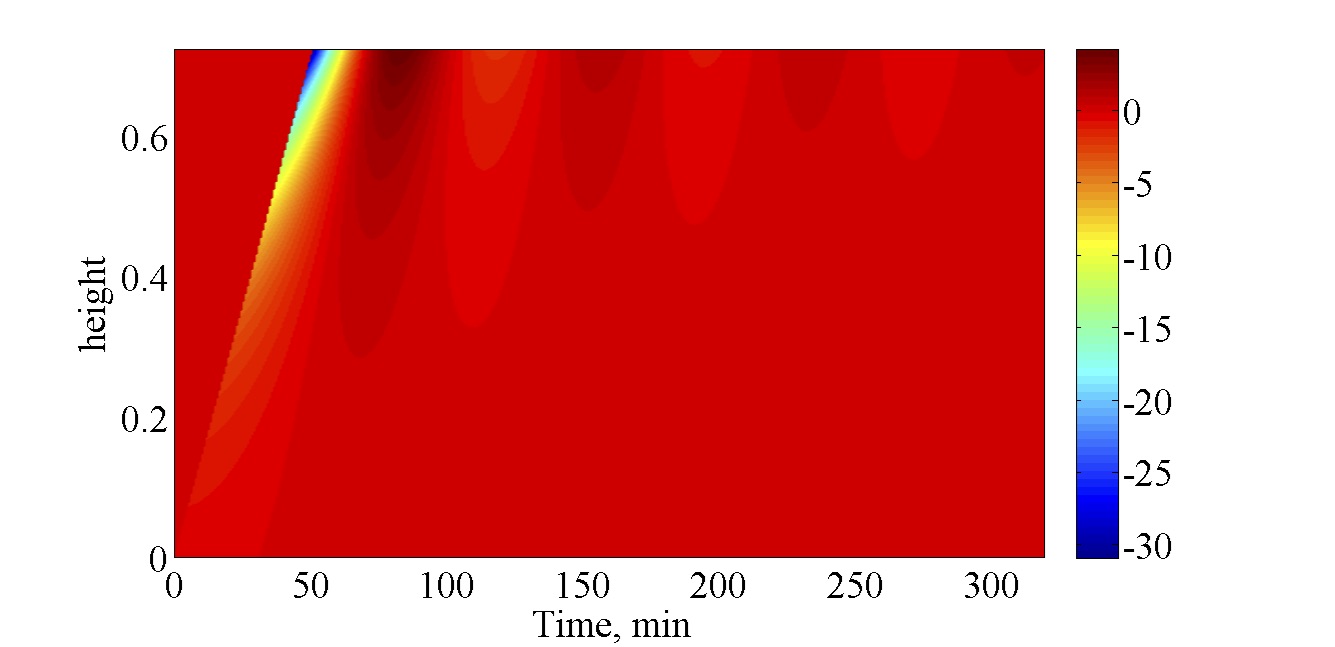}
\caption{Dynamics of solar corona after the propagation of a velocity pulse. The vertical axis shows the height above the coronal base normalized by $R_0$.  The pulse propagates with the sound speed, while the plasma starts to oscillate behind the pulse at the acoustic cut-off period. }
\label{wake}
\end{figure}%

The solution of Eq.~(\ref{8}) is the Green function for the Klein-Gordon equation \citep{Morse1953}
\begin{equation}
\label{9}
Q(z,t)={{2\pi c_s \delta(t- z/c_s) }\over {\Omega_c}}-
{{c_s}\over {\Omega_c}}{{J_1\left [\Omega_c\sqrt{t^2-{(z/c_s)^2}}\right ]}\over {\sqrt{t^2-{(z/c_s)^2}}}}H\left [\Omega_c(t- {{z}/{c_s}})\right ],
\end{equation}
where $J_1$ is the Bessel function of first order and $H$ is the Heaviside function. The first term of Eq.~(\ref{9}) represents the pulse propagating with the acoustic speed $c_s$, while the second term represents the wake oscillating at the cut-off frequency, $\Omega_c$, which is formed behind the pulse and  decays as time progresses \citep{Rae1982,Spruit1983,Roberts2004,Zaqarashvili2008,Kuridze2009}.

The wake of vertical velocity is then expressed by the formula
\begin{equation}
\label{10}
u_z(z,t)=-{{c_s}\over {\Omega_c}}{{J_1\left [\Omega_c\sqrt{t^2-{(z/c_s)^2}}\right ]}\over {\sqrt{t^2-{(z/c_s)^2}}}}H\left [\Omega_c(t- {{z}/{c_s}})\right ]exp\left [{{\gamma g}\over {2 c^2_s }}z\right ],
\end{equation}
where $g$ is the gravitational acceleration at the coronal base, $z=0$.

\begin{figure}%
\includegraphics[width=16cm]{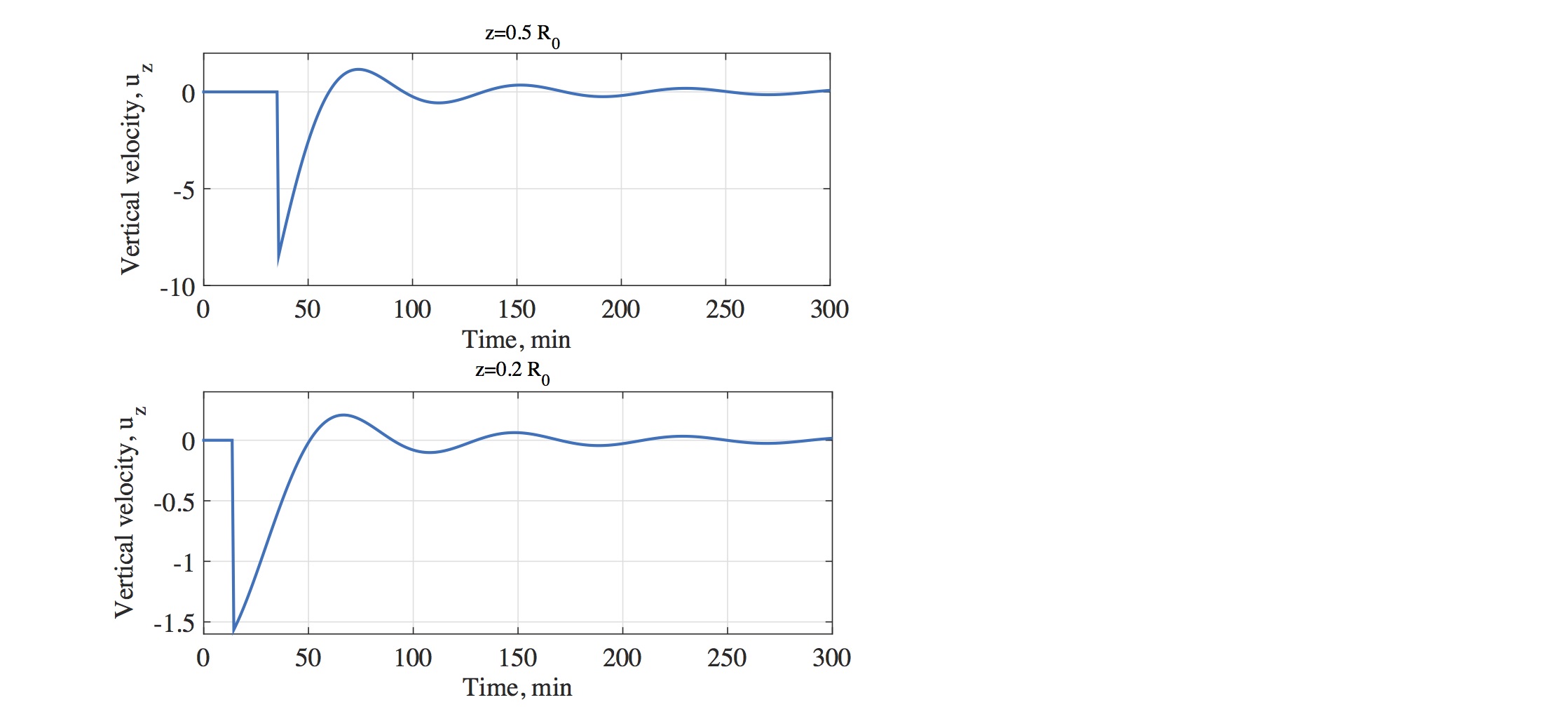}
\caption{Oscillation of wake behind the pulse at two different heights of 0.2 $R_0$ (lower panel) and 0.5 $R_0$ (upper panel) from the coronal base. The vertical velocity is normalized by the sound speed, $c_s$. The amplitude of oscillations is stronger at greater height,  but oscillations decay rapidly at each height as time progresses. }
\label{wake2}
\end{figure}

Figure~\ref{wake} displays the plot of vertical velocity from Eq.~(\ref{10}). The wake oscillating at the cut-off period, $T_c$, is formed at each height after the rapid propagation of the pulse. Figure~\ref{wake2} displays the oscillations at two different heights from the coronal base, which shows that the amplitudes of pulse and wake are increased at the greater height, due to the reduction of plasma density, but the oscillations at each height decay in time.

The plasma density has the same behaviour as the vertical velocity. If   an isothermal process ($\gamma=1$) is assumed for simplicity, then the density perturbations are governed by the same equation (Eq.~ \ref{5}) with the same consequences. Therefore, any velocity or density pulse excited near the coronal base and propagated upwards leaves an oscillating wake behind, which modulates the radio emission with the acoustic cut-off period. The mechanism of modulation is simple. Energetic electron beams are generated during the solar flare, which excites Langmuir oscillations in the solar corona. The Langmuir oscillations  emit the radio emission at observed frequencies as type IV bursts. Periodic variation in  coronal density due to the oscillating wake behind the pulse modulates the electron beam density, which influences the amplitude (and hence energy) of Langmuir oscillations and consequently the intensity of radio emission.

The scenario discussed in this section is a linear consideration. If we consider the  non-linear dynamics then the processes are governed by rebound shocks \citep{Hollweg1982}. The decrease in the equilibrium mass density with height may lead to the steepening of initial velocity or density pulse into shock. The shock propagates into
the upper corona, while the non-linear wake oscillating near the acoustic cut-off period is formed behind due to the atmospheric stratification. This non-linear wake leads to consecutive shocks. Recent numerical simulations show that the rebound shock model in the chromosphere can be responsible for the quasi-periodic occurrence of spicules and acoustic oscillations \citep{Murawski2010,Zaqarashvili2011} near the acoustic cut-off period of the chromosphere. In a similar way, the recurrent shocks may cause the density oscillations in the higher corona with the coronal acoustic cut-off period, which is much greater than that of the chromosphere.

\section{Discussion and conclusions}

Periodic density structures with length scales of hundreds to several thousands of Mm and frequencies of tens to hundreds of minutes have  regularly been observed in the solar wind at 1 AU. The similar periodic structures were recently identified by STEREO/SECCHI through COR2 images \citep{Viall2015}. They showed that  periodic structures are formed around or below 2.5 solar radii near streamers with a preferential periodicity of $\sim$ 90 minutes. The mechanism of their formation is not yet clear. However, clear magnetic features are associated with the periodic density structures when they are measured in situ \citep{Kepko2003,Viall2009-2,Kepko2016}. Often they are observed to be in pressure balance with the magnetic field around them. Therefore, a compelling idea is that these structures are related to magnetic reconnection. Indeed, the large helmet streamer blobs are thought to be related either to  full disconnection or to some form  of  interchange  reconnection. The Separatrix Web (S-Web) model combines relevant aspects of dynamic interchange reconnection behaviour and rigorous topological analysis of the solar corona \citep{Antiochos2011,Linker2011,Titov2011}. The model predicts that reconnection could occur all along the S-Web corridors which connect coronal holes. Thus, plasmoids could be released not only from the streamers. \citet{Lynch2014} performed the analysis of the 3D MHD simulation of interchange reconnection model. Spatial scales of the solar wind streamer blobs associated with reconnection-generated wave activity are similar to those observed by coronagraphs, but the temporal scales of the energy accumulation needed to drive reconnection are different. Two fluid MHD simulation \citep{Endeve2003,Endeve2004} on long timescales reveal an instability of the helmet streamer, and massive plasmoids are periodically ejected into the solar wind. However, the obtained periodicities are much longer than those detected by \citet{Viall2015}. \citet{Allred2015} were much closer to the observed periodicity in their numerical simulations as they found blobs with a period of 2 hours. However, the blobs with 2-hour periods start to appear only near 6~$R_0$, while observations already show blobs at  2.5~$R_0$. Therefore, the formation of the periodicity is an open question.

Using the URAN-2 radio telescope we observed a solar type IV burst on August 1, 2011, in the frequency range of 8--32~MHz, which corresponds to heights of 1.5--2.5~$R_0$ from the solar centre according to the Baumbach-Allen coronal density model. Wavelet analysis showed that the flux of the type IV burst undergoes  oscillation with a period of about 80~min at all frequencies. This means that the density of coronal plasma was oscillating with the same period of 80~min along the distance of 1.5--2.5~$R_0$. This periodicity is similar to that of periodic density structures obtained by \citet{Viall2015}, hence they may belong to the same phenomena.

\begin{figure}%
\centering
\includegraphics[width=8cm]{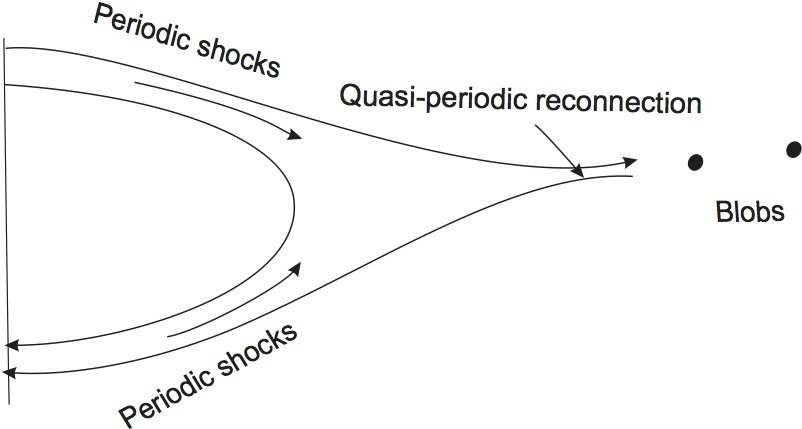}
\caption{Schematic picture of active region and excitation of quasi-periodic blobs due to recurrent shocks.}
\label{shocks}
\end{figure}

We proposed a mechanism that may explain the periodicity found in radio and white light imaging observations. The  basic physical mechanism involves the oscillations of coronal plasma at the acoustic cut-off frequency behind the pulse, which propagates in the stratified corona. We note that a similar mechanism leads to a 3 min oscillation in the solar chromosphere \citep{Fleck1991,Kuridze2009}. Any velocity or density pulse excited near the coronal base and propagated upwards leaves an oscillating wake behind, due to the density stratification. In the case of significant decrease in plasma density, the oscillations may lead to recurrent (or rebound) shocks which propagate upwards with the cut-off periodicity. These quasi-periodic shocks modulate the plasma density and consequently the amplitude of Langmuir waves, which finally lead to the quasi-periodic variations of radio emission at higher frequencies, i.e. close to the coronal base. However, the shocks propagate upward along magnetic field lines of active region and bump the starting point of helmet streamers where the field lines start to open (Figure~\ref{shocks}). The quasi-periodic shocks may trigger the quasi-periodic magnetic reconnection where the opposite field lines  merge. This reconnection will generate quasi-periodic plasma blobs which leads to periodic density structures observed in helmet streamers.    

The cut-off frequency of stratified atmosphere depends on plasma temperature. The coronal temperature of $10^6$~K leads to a periodicity of 80~min, while the temperature of $1.3\cdot10^6$~K yields a period of about 90~min. Therefore, the realistic coronal temperature may lead to the periodicity found in the radio emission of the type IV burst ($\sim$80~min) and in imaging observations ($\sim$90~min). There are two reasons for the small difference between the periods. First, Eq.~(\ref{6}) shows that the cut-off frequency decreases with height; therefore, the longer period of imaging observations may indicate a higher formation height. Second, the imaging observations may correspond to a hotter area than that of radio emission; therefore, it displays a longer periodicity.

The oscillation of the solar corona at the acoustic cut-off frequency has important consequences, if confirmed. First, the formation of quasi-periodic density structures can be explained, which is one of the current hot topics in  solar physics. Second, the oscillations can be used to estimate  plasma temperatures in the upper corona where remote sensing observations are not achievable. Third, the periodicity can be searched in the radio emission of other stars, which can give the opportunity to estimate their coronal temperatures.

More observations at different spectral bands (also with ALMA; \citealt{Wedemeyer2016}) and analytical/numerical modelling are needed to study the phenomenon in detail. 

\begin{acknowledgements}
The work was supported by the FP7-PEOPLE-2010-IRSES-269299 project  SOLSPANET. The work of TZ was supported by the Austrian ``Fonds zur F\"{o}rderung der wissenschaftlichen Forschung'' (FWF) (projects P25640-N27 and P26181-N27) and by the Georgian Shota Rustaveli National Science Foundation (projects DI-2016-17 and 217146). MP has been partly supported by the FWF project  P23762-N16.
\end{acknowledgements}


\bibliographystyle{aa}
\bibliography{manuscript29218}

\end{document}